\overfullrule=0pt
\input harvmac

\lref\Aisaka{
  Y.~Aisaka and Y.~Kazama,
  ``Origin of pure spinor superstring,''
JHEP {\bf 0505}, 046 (2005).
[hep-th/0502208].
}

\lref\BerkovitsGH{
  N.~Berkovits,
  ``Pure spinors, twistors, and emergent supersymmetry,''
JHEP {\bf 1212}, 006 (2012).
[arXiv:1105.1147 [hep-th]].
}

\lref\Matone{
  M.~Matone, L.~Mazzucato, I.~Oda, D.~Sorokin and M.~Tonin,
  ``The Superembedding origin of the Berkovits pure spinor covariant quantization of superstrings,''
Nucl.\ Phys.\ B {\bf 639}, 182 (2002).
[hep-th/0206104].
}

\lref\Bnon{
  N.~Berkovits,
  ``Pure spinor formalism as an N=2 topological string,''
JHEP {\bf 0510}, 089 (2005).
[hep-th/0509120].
}

\lref\Bpure{
  N.~Berkovits,
  ``Super Poincare covariant quantization of the superstring,''
JHEP {\bf 0004}, 018 (2000).
[hep-th/0001035].
}

\lref\Goddard{
  L.~Dolan and P.~Goddard,
  ``Tree and Loop Amplitudes in Open Twistor String Theory,''
JHEP {\bf 0706}, 005 (2007).
[hep-th/0703054].
}

\lref\Mason{
  L.~Mason and D.~Skinner,
 ``Ambitwistor strings and the scattering equations,''
[arXiv:1311.2564 [hep-th]].
}

\lref\MasonZ{
  L.~J.~Mason and D.~Skinner,
  ``Heterotic twistor-string theory,''
Nucl.\ Phys.\ B {\bf 795}, 105 (2008).
[arXiv:0708.2276 [hep-th]].
}

\lref\BerkovitsBY{
  N.~Berkovits,
  ``Ten-Dimensional Super-Twistors and Super-Yang-Mills,''
JHEP {\bf 1004}, 067 (2010).
[arXiv:0910.1684 [hep-th]].
}
\lref\BerkovitsHG{
  N.~Berkovits,
  ``An Alternative string theory in twistor space for N=4 superYang-Mills,''
Phys.\ Rev.\ Lett.\  {\bf 93}, 011601 (2004).
[hep-th/0402045].
}

\lref\Cachazo{
  F.~Cachazo, S.~He and E.~Y.~Yuan,
  ``Scattering of Massless Particles in Arbitrary Dimension,''
[arXiv:1307.2199 [hep-th]].
}

\lref\Donagi{
  R.~Donagi and E.~Witten,
``Supermoduli Space Is Not Projected,''
[arXiv:1304.7798 [hep-th]]\semi
  E.~Witten,
  ``More On Superstring Perturbation Theory,''
[arXiv:1304.2832 [hep-th]]\semi
  E.~Witten,
  ``Superstring Perturbation Theory Revisited,''
[arXiv:1209.5461 [hep-th]].
}

\lref\explaining {
N.~Berkovits,
``Explaining the Pure Spinor Formalism for the Superstring,''
JHEP {\bf 0801}, 065 (2008).
[arXiv:0712.0324 [hep-th]].
}

\lref\Berkovitssuperp{
  N.~Berkovits,
  ``Covariant quantization of the superparticle using pure spinors,''
JHEP {\bf 0109}, 016 (2001).
[hep-th/0105050].
}

\lref\osv{
  O.~Chandia,
  ``The b Ghost of the Pure Spinor Formalism is Nilpotent,''
Phys.\ Lett.\ B {\bf 695}, 312 (2011).
[arXiv:1008.1778 [hep-th]].
}

\lref\mafra{
  C.~R.~Mafra,
  ``Towards Field Theory Amplitudes From the Cohomology of Pure Spinor Superspace,''
JHEP {\bf 1011}, 096 (2010).
[arXiv:1007.3639 [hep-th]]\semi
  C.~R.~Mafra, O.~Schlotterer, S.~Stieberger and D.~Tsimpis,
  ``A recursive method for SYM n-point tree amplitudes,''
Phys.\ Rev.\ D {\bf 83}, 126012 (2011).
[arXiv:1012.3981 [hep-th]]\semi
  C.~R.~Mafra, O.~Schlotterer and S.~Stieberger,
  ``Complete N-Point Superstring Disk Amplitude I. Pure Spinor Computation,''
Nucl.\ Phys.\ B {\bf 873}, 419 (2013).
[arXiv:1106.2645 [hep-th]].
}

\lref\rennan{
  R.~L.~Jusinskas,
  ``Nilpotency of the b ghost in the non minimal pure spinor formalism,''
[arXiv:1303.3966 [hep-th]].
}

\lref\BeisertJR{
  N.~Beisert, C.~Ahn, L.~F.~Alday, Z.~Bajnok, J.~M.~Drummond, L.~Freyhult, N.~Gromov and R.~A.~Janik {\it et al.},
  ``Review of AdS/CFT Integrability: An Overview,''
Lett.\ Math.\ Phys.\  {\bf 99}, 3 (2012).
[arXiv:1012.3982 [hep-th]].
}

\lref\BerkovitsVI{
  N.~Berkovits and N.~Nekrasov,
  ``Multiloop superstring amplitudes from non-minimal pure spinor formalism,''
JHEP {\bf 0612}, 029 (2006).
[hep-th/0609012].
}

\lref\SiegelYD{
  W.~Siegel,
[arXiv:1005.2317 [hep-th]].
}

\def\bar{\overline}

\def\a{{\alpha}}

\def\ah{{\widehat \a}}

\def\ww{{\widehat w}}

\def\lh{{\widehat \lambda}}

\def\wp{{\widehat p}}
\def\wd{{\widehat d}}

\def\l{{\lambda}}

\def\b{{\beta}}
\def\bh{{\widehat\beta}}

\def\g{{\gamma}}

\def\d{{\delta}}

\def\L{{\Lambda}}

\def\half{{1\over 2}}
\def\p{{\partial}}

\def\dt{{\partial\over{\partial\tau}}}

\def\t{{\theta}}

\def\dt{{\partial_\tau}}

\def\th{{\widehat\theta}}

\Title{\vbox{\baselineskip12pt
\hbox{ICTP-SAIFR/2013-13 }}}
{{\vbox{\centerline{Infinite Tension Limit}
\smallskip
\centerline{of the Pure Spinor Superstring}}} }
\bigskip\centerline{Nathan Berkovits\foot{e-mail: nberkovi@ift.unesp.br}}
\bigskip
\centerline{\it ICTP South American Institute for Fundamental Research}
\centerline{\it Instituto de F\'\i sica Te\'orica, UNESP - Univ. 
Estadual Paulista }
\centerline{\it Rua Dr. Bento T. Ferraz 271, 01140-070, S\~ao Paulo, SP, Brasil}
\bigskip

\vskip .3in

Mason and Skinner recently constructed a chiral infinite tension limit of the Ramond-Neveu-Schwarz superstring which was shown to compute the Cachazo-He-Yuan formulae for tree-level d=10 Yang-Mills amplitudes and the NS-NS sector of tree-level d=10 supergravity amplitudes. In this letter, their chiral infinite tension limit is generalized to the pure spinor superstring which computes a d=10 superspace version of the Cachazo-He-Yuan formulae for tree-level d=10 super-Yang-Mills and supergravity amplitudes.
\vskip .3in

\Date {November 2013}

\newsec{Introduction}

Although the d=10 N=1 and N=2 superparticle describes massless 
d=10 super-Yang-Mills and supergravity states, it is 
complicated to use the worldline formulation of the superparticle to compute
d=10 super-Yang-Mills and supergravity scattering amplitudes even at tree-level.
One can of course compute these amplitudes by first computing the full 
superstring tree-level amplitudes and then taking the infinite tension limit where $\a' \to 0$, but it would be nice to have a formalism which directly computed the d=10 massless scattering amplitudes.\foot{An efficient recursive method for computing these d=10 massless
amplitudes uses the pure spinor BRST cohomology techniques developed by
Mafra and collaborators in \mafra. It would be very
interesting to relate their pure spinor recursive method with the amplitudes computed by the
infinite tension pure spinor superstring. }
These massless tree-level amplitudes can be expressed in an elegant form using the 
results of Cachazo-He-Yuan \Cachazo\ which generalize the d=4 twistor-inspired formulae to arbitrary spacetime dimension.

In a recent paper \Mason, Mason and Skinner showed how to construct a chiral infinite tension
limit of the Ramond-Neveu-Schwarz (RNS) superstring which directly computes these
massless tree-level amplitudes. For the Type II version of their superstring, they
verified for d=10 supergravity states in the NS-NS sector 
that the Cachazo-He-Yuan formulae are correctly reproduced by their tree-level superstring scattering amplitudes. For the heterotic version of their superstring, the 
Cachazo-He-Yuan formulae for d=10 Yang-Mills amplitudes are correctly reproduced but not the supergravity amplitudes. They also conjectured
that their construction could be generalized to the Green-Schwarz and pure spinor formalisms
of the superstring.

In this paper, the pure spinor generalization of their infinite tension limit of the superstring will be constructed and will be argued to compute all d=10 tree-level super-Yang-Mills and supergravity
amplitudes. The construction mirrors the Mason-Skinner approach and starts with the pure spinor versions of the d=10 N=1 and N=2 superparticle action \Berkovitssuperp. After replacing all worldline derivatives with antiholomorphic worldsheet derivatives, one obtains a conformally invariant worldsheet action which can be used to compute tree-level scattering amplitudes. The BRST operator and unintegrated vertex operators are the same as in the superparticle, and the integrated vertex operators are proportional to $\bar\d (k \cdot P)$ as in the Mason-Skinner integrated vertex operators. 

Using the relation of pure spinors and RNS fermions, it is easy to verify that the RNS and pure spinor tree-level amplitude prescriptions agree for external Yang-Mills states and for external supergravity states in the NS-NS sector. Of course, the advantage of the pure spinor formalism
is that spacetime-supersymmetry is manifest so one automatically obtains the super-Yang-Mills and supergravity amplitudes in all other sectors. Furthermore, the super-Yang-Mills and supergravity amplitudes are expessed as d=10 superspace versions of the Cachazo-He-Yun formulae which generalize the d=4 supertwistor formulae and might be useful for understanding the relation of pure spinors with
d=10 supertwistors \BerkovitsBY.

In principle, one could attempt to use this infinite tension limit of the superstring to compute
d=10 super-Yang-Mills and supergravity loop amplitudes. However, there are two reasons to suspect this will be difficult. Firstly,
d=10 super-Yang-Mills and supergravity are not consistent quantum theories because of ultraviolet divergences. Secondly, despite
the fact that the string theory describes supergravity states, the dependence on only
holomorphic worldsheet variables makes
it resembles an open string theory similar to the open twistor string theory of \BerkovitsHG.
Although there exists a closed string reformulation of the open twistor string in \MasonZ, it is unclear how to compute loop amplitudes in this type of string theory, for example, if one should use annulus worldsheets \Goddard\ or torus worldsheets.

\newsec{Infinite Tension Pure Spinor Superstring}

\subsec{Superparticle}

The d=10 N=2 superparticle in the pure spinor formalism is described by the
worldline action \Berkovitssuperp
\eqn\superptwo{
S = \int d\tau ( P_m \dt x^m + p_\a \dt \t^a + \wp_\ah \dt \th^\ah + w_\a \dt \l^\a +
\ww_\ah \dt \lh^\ah ) }
where $m=0$ to 9 are vector indices, $\a = 1$ to 16 and $\ah =1$ to 16 denote spinors of opposite chirality for the N=2A
superparticle and spinors of the same chirality for the N=2B superparticle, $(x^m,\t^\a,\th^\ah)$ are the usual variables of N=2 d=10 superspace, $(P_m, p_\a, \wp_\ah)$ are their conjugate
momenta, $\l^\a$ and $\lh^\ah$ are bosonic pure spinor variables satisfying
\eqn\pure{\l\g^m \l =0, \quad \lh\g^m\lh =0, }
and $w_\a$ and $\ww_\ah$ are their conjugate momenta which are defined up to the gauge
transformation
\eqn\pureg{\d w_\a = \L_m (\g^m\l)_\a, \quad \d \ww_\ah = \widehat\L_m (\g^m\lh)_\ah.}

Physical states are defined as ghost-number $(1,1)$ states in the cohomology of the BRST
operator 
\eqn\brsttwo{ Q = \l^\a d_\a + \lh^\ah \wd_\ah} 
where $\l^\a$ and $\lh^\ah$ carry ghost-number $(1,0)$ and $(0,1)$, and
$d_\a$ and $\wd_\ah$ are the fermionic Green-Schwarz constraints defined by
\eqn\GSc{d_\a = p_\a + \half P_m (\g^m \t)_\a, \quad \wd_\ah = \wp_\ah +\half P^m (\g^m \th)_\ah}
which satisfy the anticommutation relations
\eqn\anticom{
\{d_\a, d_\b\} = \g^m_{\a\b} P_m, \quad \{\wd_\ah, \wd_\bh\} = \g^m_{\ah\bh} P_m.}
It is easy to verify that $Q^2 =0$ using the constraints of \pure\ and the anticommutation relations of \anticom. 

The vertex operator for the N=2 superparticle is
\eqn\vertextwo{V = \l^\a \lh^\ah A_{\a\ah} (x,\t,\th) = 
e^{i k^m x_m} \l^\a A_\a(\t)\,  \lh^\ah \widehat A_\ah(\th)}
where the d=10 N=2 superfield $A_{\a\ah}(x,\t,\th)$ has been written in momentum space
and decomposed into the product of two N=1 superfields $A_\a(\t)$ and $\widehat A_\ah(\th)$.
The equation of motion $QV=0$ implies that $k^m k_m =0$ and that
\eqn\eom{ (\g_{mnpqr})^{\a\b} D_\a A_\b =0, \quad
 (\g_{mnpqr})^{\ah\bh} \widehat D_\ah \widehat A_\bh = 0}
 where $D_\a = {\p\over{\p\t^\a}} +\half k^m (\g_m\t)_\a$ and
 $\widehat D_\ah = {\p\over{\p\th^\ah}} +\half k^m (\g_m\th)_\ah$ are the
 N=2 d=10 supersymmetric derivatives. And the
 gauge invariance $\d V = Q\L$ implies that $A_\a$ and $\widehat A_\ah$ are defined
 up to the gauge transformations $\d A_\a = D_\a \L(\t)$ and 
 $\d \widehat A_\ah = \widehat D_\ah \widehat \L(\th)$.
 
In components, $A_\a(\t)$ and $\widehat A_\ah (\th)$ can be gauge-fixed onshell to the form 
\eqn\defA{ A_\a =  \half a_m (\g^m\t)_\a + {1\over 3} \xi^\b (\g^m \t)_\a (\g_m \t)_\b +  ... , }
\eqn\defAh{ \widehat A_\ah =\half  \widehat a_m (\g^m\th)_\ah +{1\over 3} \widehat\xi^\bh (\g^m \th)_\ah (\g_m \th)_\bh + ... , }
where $a_m$ and $\widehat a_m$ are vector polarizations satisfying $k^m a_m = k^m \widehat a_m=0$, $\xi^\b$ and $\widehat\xi^\bh$
are spinor polarizations satisfying
$k_m \g^m_{\a\b} \xi^\b = k_m \g^m_{\ah\bh}\widehat \xi^\bh=0$, and $...$ denotes higher-order terms in $\t^\a$ and $\th^\ah$ which
are related to the lower-order terms by BRST invariance. The polarizations of the Type II supergravity fields are expressed in
terms of these unhatted and hatted super-Yang-Mills polarizations in the usual way. The polarization of the NS-NS states $g_{mn} + b_{mn} + \eta_{mn} \phi$ is $a_{m} \widehat a_{n}$, the polarization of the
R-NS gravitino and dilatino $\chi_m^\a + \g_m^{\a\b} \rho_\b$ is $\xi^\a \widehat a_m$, 
the polarization of the
NS-R gravitino and dilatino $\widehat\chi_m^\ah + \g_m^{\ah\bh} \widehat\rho_\bh$ is $a_m \widehat \xi^\ah$, and the polarization of the R-R bispinor field-strength $F^{\a\bh}$ is
$\xi^\a \widehat\xi^\bh$.

\subsec{Type II superstring}

The generalization to the infinite tension limit of the superstring is obtained by simply
replacing worldline $\dt$ derivatives in the superparticle action of \superptwo\
with antiholomorphic $\bar\p$ worldsheet derivatives. So the infinite tension superstring action is
\eqn\superstwo{
S = \int dz d\bar z ( P_m \bar\p x^m + p_\a \bar\p \t^a + \wp_\ah \bar\p \th^\ah + w_\a \bar\p \l^\a +
\ww_\ah \bar\p \lh^\ah ) }
where all conjugate momenta variables carry conformal weight $(1,0)$. Furthermore,
the BRST operator is the same as in \brsttwo, namely
\eqn\brststwo{Q = \int dz (\l^\a d_\a + \lh^\ah \wd_\ah)}
where $d_\a$ and $\wd_\ah$ are defined in \GSc.
The left-moving stress tensor has vanishing central charge since $(x^m, P_m)$ contribute $+20$, $(p_\a,\t^\a)$ and $(\wp_\ah, \th^\ah)$ contribute $-64$, and
$(\l^\a, w_\a)$ and $(\lh^\ah, \ww_\ah)$ contribute $+44$.

To compute $N$-point tree-level scattering amplitudes using the pure spinor formalism, one
needs both unintegrated vertex operators $V$ of conformal weight $(0,0)$ and
ghost-number $(1,1)$ and integrated vertex operators $U$ of conformal weight $(1,1)$ and
ghost-number $(0,0)$. The tree-level amplitude prescription is 
\eqn\amp{ A = \langle V(z_1) V(z_2) V(z_3) 
\int d^2 z_4 U(z_4) ... \int d^2 z_N U(z_N) \rangle}
where the zero-mode measure factor is defined by
\eqn\zeromode{ \langle (\l\g^m\t)(\l\g^n\t)(\l\g^p\t)(\t\g_{mnp}\t)~(\lh\g^q\th)(\lh\g^r\th)(\lh\g^s\th)(\th\g_{qrs}\th) \rangle =1.}

The unintegrated vertex operator $V(z)$ will be defined to be the same
as the superparticle vertex operator of \vertextwo. And the integrated vertex operator
$U(z)$ will be defined to be
\eqn\defUs{ U(z) = e^{ik_m x^m} \bar\d(k^n P_n) 
[ P^m A_m(\t) + d_\a W^\a(\t) +{1\over 4} (\l\g_{mn} w) F^{mn}(\t) ]}
$$
[ P^m \widehat A_m(\th) + \wd_\ah \widehat W^\ah(\t) + {1\over 4}(\lh\g_{mn} \ww) \widehat F^{mn}(\th) ]$$
where $[A_m, W^\a, F^{mn}]$
are superfields defined in terms of $A_\a(\t)$ as
\eqn\Adefs{A_m ={1\over 8} \g_m^{\a\b} D_\a A_\b, \quad W^\a =-{1\over {10}} \g_m^{\a\b} (k_m A_\b - D_\b A_m),
\quad F_{mn} ={1\over 8} (\g_{mn})_\a{}^\b D_\b W^\a, }
$[\widehat A_m, \widehat W^\ah, \widehat F^{mn}]$
are superfields defined in terms of $\widehat A_\ah(\th)$ as
\eqn\Ahdefs{\widehat A_m = {1\over 8}\g_m^{\ah\bh} \widehat D_\ah \widehat A_\bh, \quad \widehat W^\ah =-{1\over {10}} \g_m^{\ah\bh} (k_m \widehat A_\bh - \widehat D_\bh \widehat A_m),
\quad\widehat F_{mn} ={1\over 8}  (\g_{mn})_\ah{}^\bh \widehat D_\bh \widehat  W^\ah, }
and $\bar\d(k^n P_n)$ is an operator of conformal weight $(-1,1)$ defined in the same manner as in the
integrated vertex operators of Mason and Skinner \Mason.

Note that $U(z)$ of \defUs\ is manifestly spacetime supersymmetric and is gauge-invariant
under 
\eqn\gaugett{\d A_\a = D_\a \Lambda,\quad \d A_m = k_m \Lambda, \quad 
\d \widehat A_\ah = \widehat D_\ah \widehat\Lambda, \quad
\d \widehat A_m = k_m \widehat\Lambda}
because
of the delta function $\bar\d(k^n P_n)$. Furthermore, one can verify that $QU=0$ using
the definitions of \Adefs\ and \Ahdefs\ and the fact that $k_m P^m =0$. One might be surprised
that $U$ does not involve the terms $\p\t^\a A_\a(\t)$ or $\p\th^\ah \widehat A_\ah(\th)$
which are present in the usual integrated supergravity vertex operator of the pure spinor
formalism. However, note that these terms vanish in the superparticle (since $\dt \t^\a =
\dt \th^\ah =0$ using equations of motion) and they would violate gauge invariance and
BRST invariance of $U$ because, unlike in the usual case, $QU=0$ instead of $QU = \p V$.

\subsec{Scattering amplitudes}

When the external states are in the NS-NS sector, it is straightforward to show that the
pure spinor tree-level amplitude prescription of \amp\ is equivalent to the tree-level amplitude prescription using the RNS formalism of \Mason. The equivalence proof is similar to
the proof in the standard pure spinor formalism and uses the fact that
the integrated NS-NS vertex operator of \defUs\ is
\eqn\NSNS{U =  
e^{ik_m x^m} \bar\d(k^n P_n) 
[ P^m a_m +\half k_m a_n (\t\g_{mn} p + \l \g^{mn} w) + ... ]}
$$
[ P^m \widehat a_m +\half k_m \widehat a_n (\th\g_{mn} \wp + \lh \g^{mn} \ww) + ... ]$$
where $a_m$ and $\widehat a_m$ are the unhatted and hatted polarizations of the NS-NS state
$g_{mn} + b_{mn} + \eta_{mn} \phi$, and
$...$ denotes terms which are higher-order in $\t^\a$ or $\th^\ah$. By $(p_\a,\t^\a)$ and
$(\wp_\ah, \th^\ah)$ charge conservation, one can easily verify that these higher-order terms
in $\t^\a$ and $\th^\ah$ cannot contribute to the scattering amplitude. After dropping these higher-order terms,
the vertex operator of \NSNS\ is identical to the vertex operator in the RNS formalism of \Mason\
except that the pure spinor Lorentz currents $\half(\t\g_{mn} p + \l \g^{mn} w)$ and $\half(\th\g_{mn} \wp + \lh \g^{mn} \ww)$ are replaced with the RNS Lorentz currents $\psi_m\psi_n$ and $\widehat\psi_m\widehat\psi_n$. Since these pure spinor and RNS Lorentz currents generate 
SO(9,1) Kac-Moody algebras of the same level \Bpure, the OPE's of the pure spinor vertex operators
of \NSNS\ are identical to the OPE's of the RNS vertex operators of \Mason. 

Furthermore, note that the zero mode measure factor of \zeromode\ correctly
reproduces the three-point supergravity amplitude when all three vertex operators
are unintegrated (for the same reason as in the usual pure spinor superstring). Combining
this fact with the equivalent OPE's of the integrated NS-NS vertex operators implies that the
pure spinor ampitude prescription of \amp\ agrees with the RNS amplitude prescription when all external states are NS-NS, which
was shown by Mason and Skinner in
\Mason\ to correctly reproduce the Cachazo-He-Yuan formulae of \Cachazo.

In addition, the pure spinor amplitude prescription of \amp\ automatically provides a manifestly supersymmetric
generalization of the Cachazo-He-Yuan formulae which describes in d=10 superspace the
tree-level amplitudes of d=10 N=2 supergravity. It would be very interesting to work out the properties of this supersymmetric formula.

\subsec{Heterotic superstring}

Finally, one can easily generalize these results to d=10 super-Yang-Mills by
constructing the pure spinor analog of the Mason-Skinner heterotic superstring of \Mason.
In this case, the hatted variables of \superstwo\ are replaced with a holomorphic current algebra contributing central charge $+16$ and a set of $(b,c)$ Virasoro ghosts contributing
central charge $-26$. So the infinite tension limit of
the pure spinor heterotic superstring action is
\eqn\supershet{
S = \int dz d\bar z ( P_m \bar\p x^m + p_\a \bar\p \t^a + w_\a \bar\p \l^\a +
\bar b \bar\p c)  + S_C}
where $S_C$ is the worldsheet action for the current algebra.

The BRST operator for this heterotic superstring is
\eqn\brsth{ Q = \int dz (\l^\a d_\a + c (P_m \p x^m + p_\a \p \t^\a + w_\a \p\l^\a + T_C)  + b c \p c )}
where $T_{C}$ is the $c=16$ stress-tensor of the current algebra. And the tree-level amplitude
prescription is
\eqn\amph{ A = \langle V(z_1) V(z_2) V(z_3) 
\int d^2 z_4 U(z_4) ... \int d^2 z_N U(z_N) \rangle}
where $V$ and $U$ are unintegrated and integrated vertex operators in the BRST
cohomology, and the zero-mode measure factor is defined by
\eqn\zeromode{ \langle (\l\g^m\t)(\l\g^n\t)(\l\g^p\t)(\t\g_{mnp}\t)~c \p c \p^2 c \rangle =1.}

The unintegrated vertex operator $V$ describing the super-Yang-Mills multiplet is
\eqn\uninth{ V =  e^{ik_m x^m}  c \l^\a A^I_\a (\t)  J^I }
where $A^I_\a(\t)$ is the super-Yang-Mills spinor gauge field, $I$ denotes the adjoint representation of the gauge group,
 and $J^I$ are the holomorphic currents of conformal weight $(1,0)$.
$QV=0$ implies that $k^m k_m=0$ and $\g_{mnpqr}^{\a\b} D_\a A^I_\b =0$, and
$\d A_\a^I = D_\a \Lambda^I$ under the gauge transformation $\d V = Q (c \L^I J^I)$.
As in \defA, $A_\a^I$ can be gauge-fixed onshell to
\eqn\defAA{ A_\a^I =  \half a_m^I (\g^m\t)_\a +{1\over 3} \xi^{I\b} (\g^m \t)_\a (\g_m \t)_\b +  ...  }
where $a_m^I$ and $\xi^{I\a}$ are the gluon and gluino polarizations satisfying
$k^m a_m^I = k_m \g^m_{\a\b} \xi^{I\b}=0$, and $...$ are terms higher-order in $\t^\a$ which are related to the lower-order terms by BRST invariance.
Although one also can define the unintegrated vertex operator
$V = e^{i k_m x^m} c \l^\a A_{\a m}(\t) P^m$, this operator does not appear to
correctly describe supergravity \Mason.

The integrated vertex operator $U$ describing the super-Yang-Mills multiplet is defined
in analogy with \defUs\ as  
\eqn\defUsh{ U(z) = e^{ik_m x^m} \bar\d(k^n P_n) 
[ P^m A^I_m(\t) + d_\a W^{I\a}(\t) +{1\over 4} (\l\g_{mn} w) F^{Imn}(\t) ] J^I}
where $[A^I_m, W^{I\a}, F^{Imn}]$ are defined in terms of $A_\a^I(\t)$ as in \Adefs.
Gauge invariance and BRST invariance of \defUsh\ are verified as in \gaugett.

Using the same arguments as in the previous subsection, one can check that the
pure spinor tree-level amplitude prescription of \amph\ reproduces the RNS
tree-level amplitude prescription of \Mason\ when all external states are Yang-Mills
gluons. Since these d=10 Yang-Mills amplitudes were shown in \Mason\ to coincide with the
Cachazo-He-Yuan formulae of \Cachazo, the prescription of \amph\  provides
a supersymmetric generalization of these formulae to d=10 super-Yang-Mills.
Hopefully, this supersymmetric generalization of twistor-inspired formulae
will be useful for identifying the
appropriate d=10 generalization of d=4 supertwistors.

\vskip 10pt
{\bf Acknowledgements:}
I would like to thank CNPq grant 300256/94-9
and FAPESP grants 09/50639-2 and 11/11973-4 for partial financial support.

\listrefs
\end